\documentclass[journal=jacsat,manuscript=article]{achemso}
\setkeys{acs}{etalmode=truncate,maxauthors=10}
\usepackage[version=3]{mhchem} 
\usepackage{amssymb, physics}
\usepackage{xcolor}
\usepackage{rotating}
\usepackage{subcaption,hyperref}
\usepackage{graphicx}
\usepackage{float}

\newcommand\scientific[2]{#1 \cdot 10^{#2}}

\newcommand{\tocless}[2]{\bgroup\let\addcontentsline=\nocontentsline#1{#2}\egroup}

\author{Akhil Shajan}
\affiliation{Center for Computational Life Sciences, Lerner Research Institute, The Cleveland Clinic, Cleveland, Ohio 44106, United States}

\author{Danil Kaliakin}
\affiliation{Center for Computational Life Sciences, Lerner Research Institute, The Cleveland Clinic, Cleveland, Ohio 44106, United States}

\author{Fangchun Liang}
\affiliation{Center for Computational Life Sciences, Lerner Research Institute, The Cleveland Clinic, Cleveland, Ohio 44106, United States}

\author{Thaddeus Pellegrini}
\affiliation{IBM Quantum, IBM T.J. Watson Research Center, Yorktown Heights, NY 10598, United States}

\author{Hakan Doga}
\affiliation{IBM Quantum, IBM T.J. Watson Research Center, Yorktown Heights, NY 10598, United States}

\author{Subhamoy Bhowmik}
\affiliation{Center for Computational Life Sciences, Lerner Research Institute, The Cleveland Clinic, Cleveland, Ohio 44106, United States}
\alsoaffiliation{Department of Chemistry, Michigan State University, East Lansing, Michigan 48824, United States}

\author{Susanta Das}
\affiliation{Center for Computational Life Sciences, Lerner Research Institute, The Cleveland Clinic, Cleveland, Ohio 44106, United States}

\author{Antonio Mezzacapo}
\affiliation{IBM Quantum, IBM T.J. Watson Research Center, Yorktown Heights, NY 10598, United States}

\author{Mario Motta}
\affiliation{IBM Quantum, IBM T.J. Watson Research Center, Yorktown Heights, NY 10598, United States}

\author{Kenneth M. Merz Jr.}
\email{kmerz1@gmail.com}
\affiliation{Center for Computational Life Sciences, Lerner Research Institute, The Cleveland Clinic, Cleveland, Ohio 44106, United States}
\alsoaffiliation{Department of Chemistry, Michigan State University, East Lansing, Michigan 48824, United States}

\title[An \textsf{achemso} demo]
  {Molecular Quantum Computations on a Protein}

\abbreviations{IR,NMR,UV}

\SectionNumbersOn

\begin{document}

\begin{abstract}
This work presents the implementation of a fragment-based, quantum-centric supercomputing workflow for computing molecular electronic structure using quantum hardware. The workflow is applied to predict the relative energies of two conformers of the 300-atom Trp-cage miniprotein. 
The methodology employs wave function–based embedding (EWF) as the underlying fragmentation framework, in which all atoms in the system are explicitly included in the CI treatment. CI calculations for individual fragments are performed using either sample-based quantum diagonalization (SQD) for challenging fragments or full configuration interaction (FCI) for trivial fragments. To assess the accuracy of SQD for fragment CI calculations, EWF-(FCI,SQD) results are compared against EWF-MP2 and EWF-CCSD benchmarks. 
Overall, the results demonstrate that large-scale electronic configuration interaction (CI) simulations of protein systems containing hundreds or even thousands of atoms can be realized through the combined use of quantum and classical computing resources.
\end{abstract}


\section{Introduction}

Simulations of the electronic structure of molecules are one of the most promising applications of quantum computing \cite{schleich2025cracking, weidman2024quantum, alexeev2024quantum, motta2022emerging}. The recently embraced paradigm of quantum-centric computing, where classical high-performance computing (HPC) and quantum processing units (QPUs) work in concert with each other, showed that electronic structure simulations utilizing present day quantum computers can be scaled up to at least 36 molecular orbitals (MOs) and beyond 72 qubits \cite{robledo2024chemistry, shirakawa2025closedloopcalculationselectronicstructure} bringing the possibility of useful quantum computing simulations closer to reality. 

Quantum-centric computing of the electronic structure of a molecular system is currently enabled using the quantum-selected configuration interaction (QSCI) method \cite{kanno2023quantum} and its variant, the sample-based quantum diagonalization methodology (SQD) \cite{robledo2024chemistry,shirakawa2025closedloopcalculationselectronicstructure}. The SQD methodology introduced an expansion of the QSCI method through an error mitigation procedure, called configuration recovery, that allows the recovery of samples of electron configurations that violate known symmetries of the ansatz state methodology\cite{robledo2024chemistry}. The SQD methodology was further expanded with the introduction of the carryover procedure that preserves the most dominant electron configurations identified within each step of configuration recovery\cite{shirakawa2025closedloopcalculationselectronicstructure}.

QSCI and SQD methodologies are both forms of selected configuration interaction (SCI) method and the hope is these methods will offer eventual quantum advantage through efficient sampling of relevant electron configurations. Despite a recent criticism of the sampling ability of the QSCI and SQD methodologies\cite{reinholdt2025critical}, the original SQD paper demonstrated through numerical experiments demonstrated the existence of quantum circuits enabling the electronic configuration sampling within the SQD method that outperforms the fully-classical heat-bath configuration interaction method (HCI)\cite{robledo2024chemistry}. Moreover, a recent QSCI study demonstrated that the hardware runs are closely approaching the regime where quantum sampling can offer advantages over fully-classical simulations\cite{weaving2025towards}.

Quantum-centric computing of electronic structure has already been shown to be promising in applications to studies of metal complexes\cite{robledo2024chemistry,shirakawa2025closedloopcalculationselectronicstructure,nutzel2024solving}, aromatic molecules\cite{sugisaki2025hamiltonian, piccinelli2025quantum, shirai2025enhancing}, intermolecular interactions\cite{kaliakin2024accurate, sugisaki2025size}, open-shell systems\cite{sugisaki2025hamiltonian, Ieva2024}, excited states of molecular systems ~\cite{shirai2025enhancing,barison2025quantum}, systems of interest for material science\cite{barroca2025surface,nogaki2025symmetry,wray2025convergence}, electronic structure of molecules in implicit solvent\cite{kaliakin2025implicit}, quantum mechanics / molecular mechanics (QM/MM) \cite{bickley2025extending,bazayeva2025quantum}, as well as free energy with molecular dynamics (MD) in explicit solvent \cite{bazayeva2025quantum}. Moreover, the SQD method was evaluated against the W4-11 thermochemistry dataset that includes 124 total atomization, 83 bond dissociation, 20 isomerization, 505 heavy-atom transfer, and 13 nucleophilic substitution processes.  This evaluation demonstrated the ability of the SQD method to describe a diverse set of chemical environments and reactions.\cite{raisuddin2025promise}

Application of quantum-centric electronic structure simulations to problems of relevance to healthcare and life sciences inevitably requires scaling of these simulations to biomolecular systems like proteins\cite{baiardi2023quantum,pal2024future}, where applications of quantum mechanical (QM) methods were previously shown to provide high-accuracy in the prediction of disordered regions within a protein structure\cite{ufimtsev2011charge} as well as prediction of protein-ligand binding free energies\cite{mobley2017predicting}. However, despite substantial successes in scaling up QM simulations, existing classical computational resources are insufficient to carry out first-principles electronic structure simulations on protein systems by methods beyond the Hartree-Fock (HF), density functional theory (DFT), and Second-order Møller–Plesset perturbation theory (MP2)\cite{ching2021ultra,stocks2024breaking}. Recent advances in GPU acceleration further pushed the scalability of coupled cluster singles, doubles, and perturbative triples (CCSD(T)) simulations\cite{fajen2025accelerating}, as well as full configuration interaction simulations\cite{fales2020efficient}, but these simulations on their own are still far from being scalable to large protein systems. 

Configuration interaction simulations are especially challenging to scale up\cite{gao2024distributed}. In a 2013 study Yoshikawa et al.\cite{yoshikawa2013novel} demonstrated the divide-and-conquer (DC) symmetry-adapted cluster configuration interaction (SACCI)\cite{yoshikawa2013divide} simulation of photoactive yellow protein where only the chromophore and five surrounding amino-acid residues were treated with the SACCI method. The heat-bath configuration interaction (HCI) has been reported to be scaled up to only (12e,
190o),  (14e, 108o) and (48e, 42o) active spaces\cite{holmes2016heat,sharma2017semistochastic}, while the largest DMRG simulations are limited to (113e, 76o) and (63e, 58o).\cite{menczer2024parallel} Finally, the advanced formulations of the complete active space self-consistent field (CASSCF) method including the many-body expanded full configuration interaction approach\cite{greiner2024mbe} and adaptive selected configuration approach\cite{levine2020casscf} are limited to simulations of (40e,42o), (50e, 50o), and (52e,52o) active spaces.

Quantum mechanical calculations beyond the mean-field approach in large protein systems require the utilization of approximations enabling this class of large-scale simulations. Such approximations can be realized using fragment-based electronic structure methods.~\cite{kitaura1999fragment, raghavachari2015accurate, erlanson2019fragment} These algorithms are based on the decomposition of a prohibitively large system into manageable subsystems, and typically employ a quantum embedding scheme, where subsystems are treated with the higher-level method while lower-level methods are used to generate the initial mean-field object corresponding to the entire system~\cite{sun2016quantum}. Previous studies proposed the integration of quantum computing subroutines in the workflow of quantum embedding methods~\cite{bauer2016hybrid, kreula2016few, bravyi2017complexity} through feedback between subsystems and their environment based on the electronic density, the Green’s function, or the one-particle reduced density matrix.~\cite{sun2016quantum}

Early success in integration of quantum computing with fragment-based methods was achieved with the variational quantum eigensolver (VQE) method~\cite{peruzzo2014variational, motta2022emerging, bauer2020quantum} in combination with the fragment molecular orbital (FMO),~\cite{lim2024fragment} divide and conquer (DC),~\cite{yoshikawa2022quantum} many-body expansion (MBE),~\cite{ma2023multiscale} density matrix embedding theory (DMET),~\cite{rubin2016hybrid, kawashima2021optimizing, iijima2023towards, li2022toward, kirsopp2022quantum, cao2023ab, ma2023multiscale} and other wavefunction-based embedding ~\cite{vorwerk2022quantum, rossmannek2023quantum, gujarati2023quantum} methods. However, utilization of VQE hindered the scalability of these simulations leading to these simulations being performed either with classical simulators of quantum circuits or being limited to quantum hardware execution using a maximum of 12 qubits\cite{ma2023multiscale}.

In our recent study we demonstrated that the SQD method can be combined with the DMET method allowing for scaling to simulations using 32 qubits, which substantially exceeds what was reported before with VQE-based methodologies\cite{shajan2025toward}. Moreover, unfragmented systems demonstrated in this study would require 41 and 89 qubits for a chain of 18 hydrogen atoms ($H_{18}$) and cyclohexane, respectively. Our initial success with the DMET-SQD approach lead to a recent publication by Patra et al. where researchers adapted the approach originally proposed in our study to simulate drug-like molecules\cite{patra2025quantum}. Moreover, Bierman et al. demonstrated another promising fragment-based simulation with SQD using quantum bootstrap embedding\cite{bierman2025towards}.

\begin{figure}[h]
    \includegraphics[width=1\textwidth]{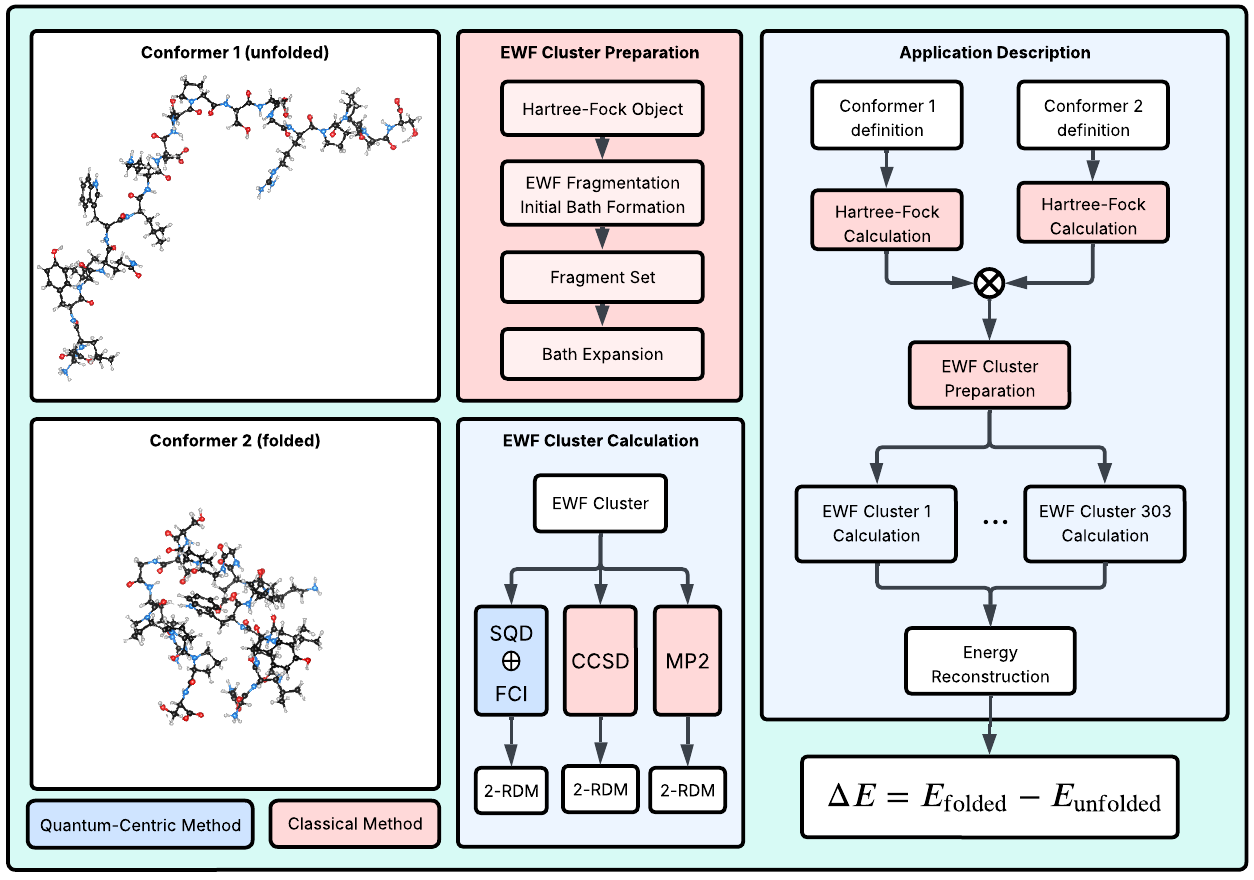} 
    \caption{Overview of the EWF-based quantum-centric workflow for Trp-Cage conformer energetics. The protocol includes Hartree–Fock reference generation, automated fragmentation and bath construction, bath orbital expansion, and high-level cluster calculations using either quantum (SQD/FCI) or classical (MP2/CCSD) solvers. Fragment energies are reconstructed to obtain total conformer energies.} 
    \label{fig:method_summary} 
\end{figure}

In the present work, we show that embedding-based quantum centric simulations can be scaled even further to realistic protein models. To achieve such scalability, we moved beyond the original DMET formulation in favor of the wavefunction-based embedding method (EWF) which represents the further expansion of the base DMET technology. The EWF formulation offers three important benefits over the DMET formulation: 1) unlike the standard DMET method it allows for more flexible extension of the bath (surrounding entangled orbitals) through MP2-based calculations; 2) it eliminates the loop over the chemical potential allowing for a more efficient non-iterative procedure which further reduces the overall computational cost of simulations\cite{nusspickel2022systematic}; 3) because of the way EWF treats the fragment plus it's bath, we can make the individual fragments as small as individual atoms, while still recovering high quality results. Figure \ref{fig:method_summary} shows the EWF workflow adapted in present study and demonstrates how we integrate SQD and FCI solvers for quantum-centric calculations as well as how we utilize CCSD and MP2 solvers for classical benchmark

The Trp-cage miniprotein provides an ideal molecular benchmark for evaluating quantum embedding techniques due to its compact size, biological relevance, and the presence of multiple energetically distinct conformers\cite{simmerling2002all}. Although only ~20 residues long (~300 atoms), Trp-cage exhibits a well-defined hydrophobic core formed by the indole side chain of tryptophan, as well as a characteristic $\alpha$-helix and poly-proline loop\cite{maruyama2023effect,ozono2021visualization}. In this work, we focus on two representative structures: the lowest-energy folded conformer, and the unfolded form\cite{simmerling2002all}. Their differing geometries lead to distinct electronic environments, including variations in charge distribution, hydrogen bonding, and steric effects. Orbital localization and fragmentation of these geometries yield EWF clusters (fragment + expanded bath) spanning 6 to 33 molecular orbitals. Smaller fragments can be treated exactly, while larger ones exhibit substantial correlation and require more advanced solvers. This wide distribution of EWF cluster sizes makes Trp-cage a stringent test for the EWF methodology and highlights the need for a scalable solver strategy. 

We first describe the EWF embedding methodology and the quantum-centric CI solver used for fragment calculations in the Methods section. The Computational Details section then summarize the software components, parameters for fragment construction, and the criteria for selecting between SQD and FCI solvers within the EWF-(FCI,SQD) workflow. We also outline the extraction of candidate electronic configurations from quantum hardware and the associated classical post-processing in SQD, and analyze the quantum and classical computational complexity of the Trp-cage simulations. The Results section compares EWF-(FCI,SQD) with fully classical approaches, and the Conclusions and Outlook highlight key advances and future directions for quantum-centric EWF methodologies.

\section{Methods}

\subsection{Embedded Wave Function (EWF)}

The Embedded Wave Function (EWF) framework \cite{Booth2022,Booth2023} builds upon the Density Matrix Embedding Theory (DMET) formulation, which uses a mean-field calculation to define localized fragments and DMET bath orbitals that reproduce the one-particle entanglement between each fragment and its environment. EWF departs from traditional DMET in how correlation is treated and how global properties are reconstructed. Instead of enforcing self-consistency through matching one-particle density matrices, each fragment-and-bath EWF cluster is solved at a correlated level, and the resulting one- and two-body density matrices are projected back into the full molecular orbital basis. These projected quantities are then combined across EWF clusters to obtain global expectation values, including the total energy.

The construction and correlation of these EWF clusters proceeds in two main steps, described in the following subsections. First, the fragments and their mean-field entanglement baths are defined using localized orbitals (Fragment and Bath Construction). Second, the minimal DMET bath is systematically expanded using correlated information to recover the relevant local excitations and improve the representation of fragment–environment entanglement (Interacting Bath Expansion).

\subsubsection{Fragment and Bath Construction}

Fragmentation in EWF employs Intrinsic Atomic Orbitals (IAOs) to define a localized and chemically meaningful basis for the occupied space. A full-system mean-field calculation provides the reference one-particle density matrix, and the associated fragment–environment entanglement is captured using the standard DMET Schmidt decomposition. This procedure yields a compact entanglement bath that exactly reproduces the mean-field hybridization for each fragment.

\subsubsection{Interacting Bath Expansion}

In the DMET literature \cite{chen2014intermediate,ma2023multiscale,rubin2016hybrid}, the fragment and its entangled surroundings are commonly referred to as the impurity. In the EWF framework, the initial DMET bath is expanded in a manner analogous to pair natural orbitals, introducing additional occupied and virtual states that systematically recover the local excitations needed for an accurate correlated fragment description and an improved representation of fragment–environment entanglement. To identify the most relevant additions to this space, the local interacting subspace from second-order Møller–Plesset perturbation theory (MP2) is costructed. Two subspace MP2 calculations are performed for each fragment: one in which the occupied excitations are restricted to the DMET impurity space, and another in which the virtual excitations are similarly restricted. The resulting correlated one- and two-body density matrices are used to form bath natural orbitals spanning the corresponding occupied and virtual environment spaces. Their occupation numbers quantify the strength of correlation-driven coupling to the fragment. Natural orbitals with occupations larger than a threshold $\eta$ (for virtual) or smaller than $2-\eta$ (for occupied) are retained and included with the fragment and original DMET bath to define the final correlated EWF impurity. Following this expansion, the combined fragment–bath unit is referred to as an \textit{EWF cluster}, consistent with the original EWF terminology \cite{Booth2022}. 

\subsection{Sample-based quantum diagonalization (SQD) and extended SQD (ext-SQD)}

Recent developments in hybrid quantum-classical algorithms have unlocked new possibilities for scaling up applications of quantum computing for chemistry. These are based on the classical processing of individual quantum samples, such as the quantum-selected configuration interaction method~\cite{kanno2023quantum}, which extended ideas from the classical configuration interaction methods to the quantum domain, and the sample-based quantum diagonalization (SQD) workflow, which allowed scaling for up to 85 qubits~\cite{robledo2024chemistry, yu2025quantum} for molecular and fermionic lattice models.
We sample candidate electron configurations, to be used within the SQD workflow, from the local unitary cluster Jastrow (LUCJ)~\cite{motta2023bridging} ansatz. Such an approach leverages the idea that appropriate quantum circuits can in principle produce an electron configuration subspace that is significantly more compact than a subspace produced with a fully-classical heuristic, while still predicting the accurate total energy of the target system. The sampling of electron configurations with the LUCJ circuit was demonstrated through numerical experiments in the original SQD paper.~\cite{robledo2024chemistry} 
The quantum-centric aspect of SQD involves using a quantum computer to generate candidate configurations, and classical high-performance computing (HPC) resources to carry out error mitigation at the level of individual samples and to perform diagonalization of a projected Hamiltonian.

After identifying a subspace with SQD, we further improve accuracy by first identifying the most relevant configurations within the obtained subspaces and then extend these configurations through single excitation operators (so-called extended-SQD~\cite{barison2025quantum}). 
The resulting subspace allows for even more extensive capture of relevant configurations and improves the quality of the resulting total energy and reduced density matrices.

\section{Computational Details}

\subsection{Classical pre-processing}

\subsubsection{Conformer definition}

In this work, we calculate the relative energy between the most energetically stable and most energetically unstable conformers of Trp-cage (henceforth called the ``folded'' and ``unfolded'' conformers), as defined in a previous study by Simmerling et al.\cite{Simmerling2002} We generated the protonated structures of these two systems using the H++~\cite{anandakrishnan2012h++, myers2006simple, gordon33h++} web server at pH = 7. During the protonation process we only add hydrogen atoms that correspond to protonation of amino acids at a given pH. However, we are not using explicit or implicit solvents, and the simulations are performed in the gas phase. Upon protonation, the overall charge of the system is +1. To date, no DMET or EWF studies have demonstrated the simulation of charged species \cite{Booth2022, Booth2023}. We believe that the charged cases represent a challenge for these methods and require further detailed study. To maintain the most biologically relevant protonation while changing the nature of the overall system to a zwitterionic form (where individual charged residues lead to an overall neutral charge of the system), we doprotonated LYS8 via removal of one of the hydrogen atoms from the NH$_3$ group, for both the folded and unfolded conformers. We chose LYS8 for modification because it is located on the surface of the protein in both conformers and it is not involved in critical non-covalent interactions. We treated the system as a closed-shell singlet spin state. 

\subsubsection{Mean-field calculations}
All mean-field calculations for the folded and unfolded Trp-cage conformers were performed using GPU4PySCF \cite{li2024introducting, wu2024enhancing} employing the restricted HF method with the STO-3G basis set (HF/STO-3G). Density fitting \cite{B000027M} was enabled throughout to reduce the overhead of the integral evaluation and maximize throughput on GPU hardware. The default PySCF auxiliary basis applied during density fitting was def2-SVP-JKFIT for the elements present in Trp-cage (H, C, N, and O). Each calculation was executed on a single NVIDIA V100S-PCIe 32 GB accelerator paired with a 2.4 GHz Intel Xeon Platinum 8260 CPU. The GPU-accelerated Hartree-Fock evaluations finished in roughly four minutes per conformer. This mean-field stage establishes the orbital and density matrix input that feeds into the subsequent fragmentation, bath construction, and correlated solver workflows.

\subsubsection{Embedded Wave Function Fragmentation}

All calculations were carried out using an IAO-based localized orbital framework to define chemically meaningful fragments and to construct the initial DMET entanglement baths. A full-system Hartree–Fock calculation was used to generate the reference one-particle density matrix, from which the IAOs and projected virtual orbitals were obtained.

Following the DMET construction, each fragment and its corresponding mean-field bath were subjected to an interacting bath expansion at the MP2 level. The MP2-derived one- and two-body density matrices were used to construct bath natural orbitals, whose occupation numbers quantify the degree of correlation-driven coupling between the fragment and its environment. In this work, a numerical threshold of $\eta=\scientific{1}{-5}$ was applied to select the most relevant bath natural orbitals. This augmentation introduces additional virtual degrees of freedom necessary to capture dynamical correlation, long-range polarization, and environment-mediated fluctuations that are not described by the minimal DMET bath.

The calculations were performed on hardware equipped with 197 GB of RAM and four CPU cores. For each conformer, the interacting bath expansion requires approximately 487 MB of disk storage. Memory usage depends on the number of bath natural orbitals retained and therefore scales with the chosen threshold. At the threshold of $\eta=\scientific{1}{-5}$, all EWF cluster calculations comfortably fit within available memory resources without the need for disk-based integral batching.

The typical wall-time for a single EWF cluster formation and MP2-level bath expansion is around 30 minutes, although this increases when tighter orbital-selection thresholds are applied or when fragments contain more atoms and therefore require larger correlated subspaces. For workflows involving multiple conformers, each fragment–bath calculation was run independently, enabling trivial parallelization across computational nodes.

\begin{figure}[h]
    \centering
    \includegraphics[width=0.8\textwidth]{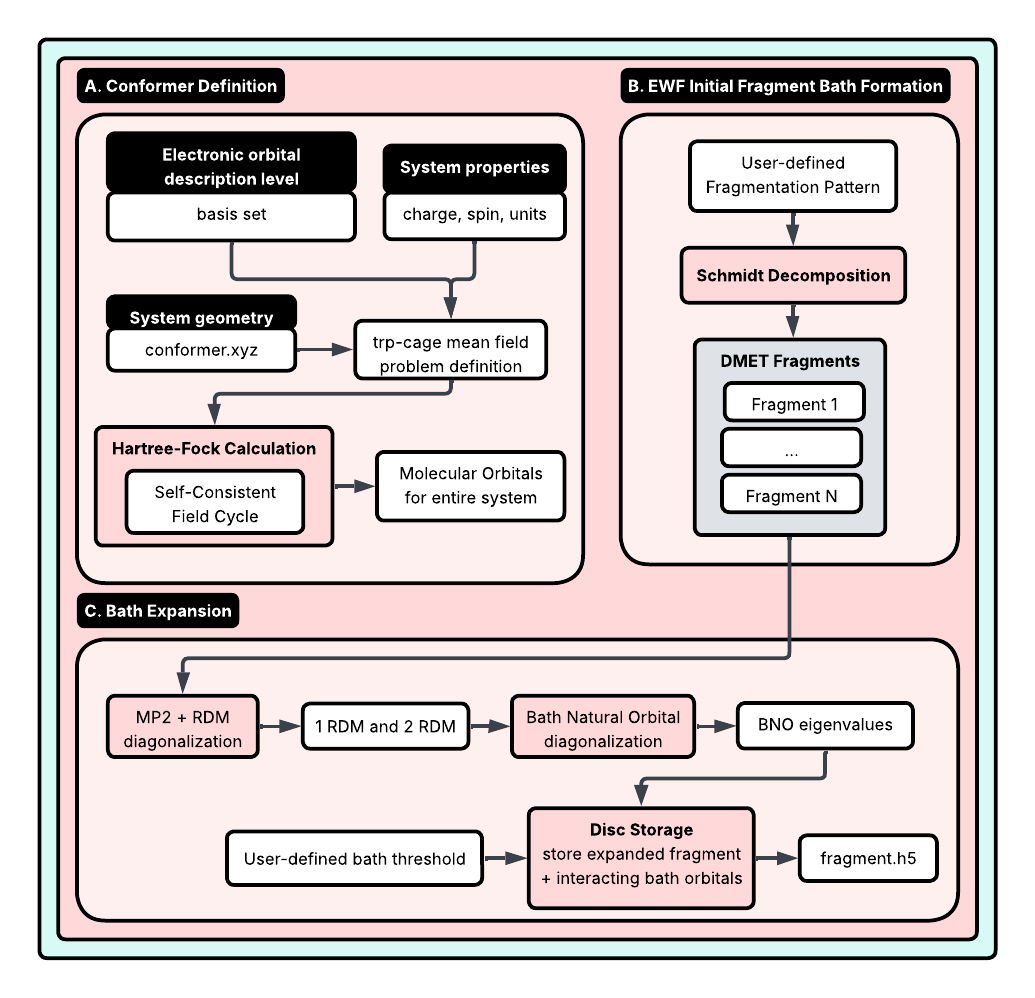}
    \caption{Schematic overview of the classical pre-processing and EWF workflow. The diagram illustrates (A) conformer definition and mean-field problem setup, (B) initial EWF fragment and DMET bath construction from the Hartree--Fock reference, and (C) MP2-level interacting bath expansion via bath natural orbitals.}
    \label{fig:application_description}
\end{figure}

An overview of the full classical pre-processing workflow, including conformer definition, mean-field calculations, initial EWF fragment–bath construction, and MP2-based bath expansion, is summarized schematically in Fig.~\ref{fig:application_description}.

\clearpage
\subsection{Post Hartree-Fock computations of EWF clusters}
\begin{figure}[h] 
    \includegraphics[width=0.8\textwidth]{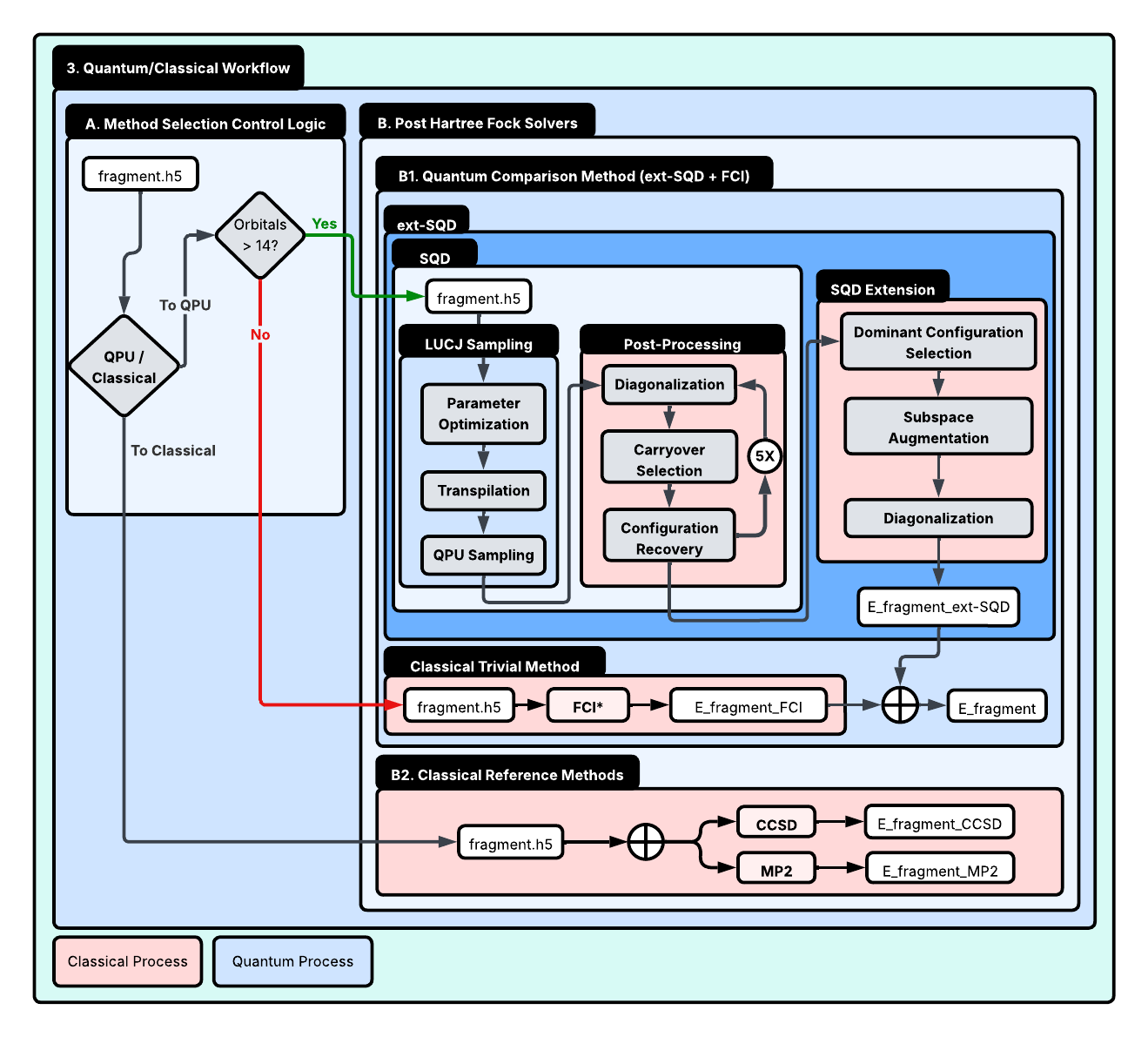} 
    \caption{Quantum--classical workflow for post--Hartree--Fock EWF cluster calculations. The diagram summarizes solver selection logic, quantum circuit execution, and classical post-processing steps used in EWF-(SQD,FCI) calculations.}    
    \label{fig:qc_workflow} 
\end{figure}

Figure~\ref{fig:qc_workflow} provides a schematic overview of the post--Hartree--Fock quantum--classical workflow used to solve EWF cluster Hamiltonians, including solver selection, quantum circuit execution, and classical post-processing.

\subsubsection{Choice of Correlated Solver for EWF cluster Spaces}
\begin{figure}
    \centering
    \includegraphics[width=\linewidth]{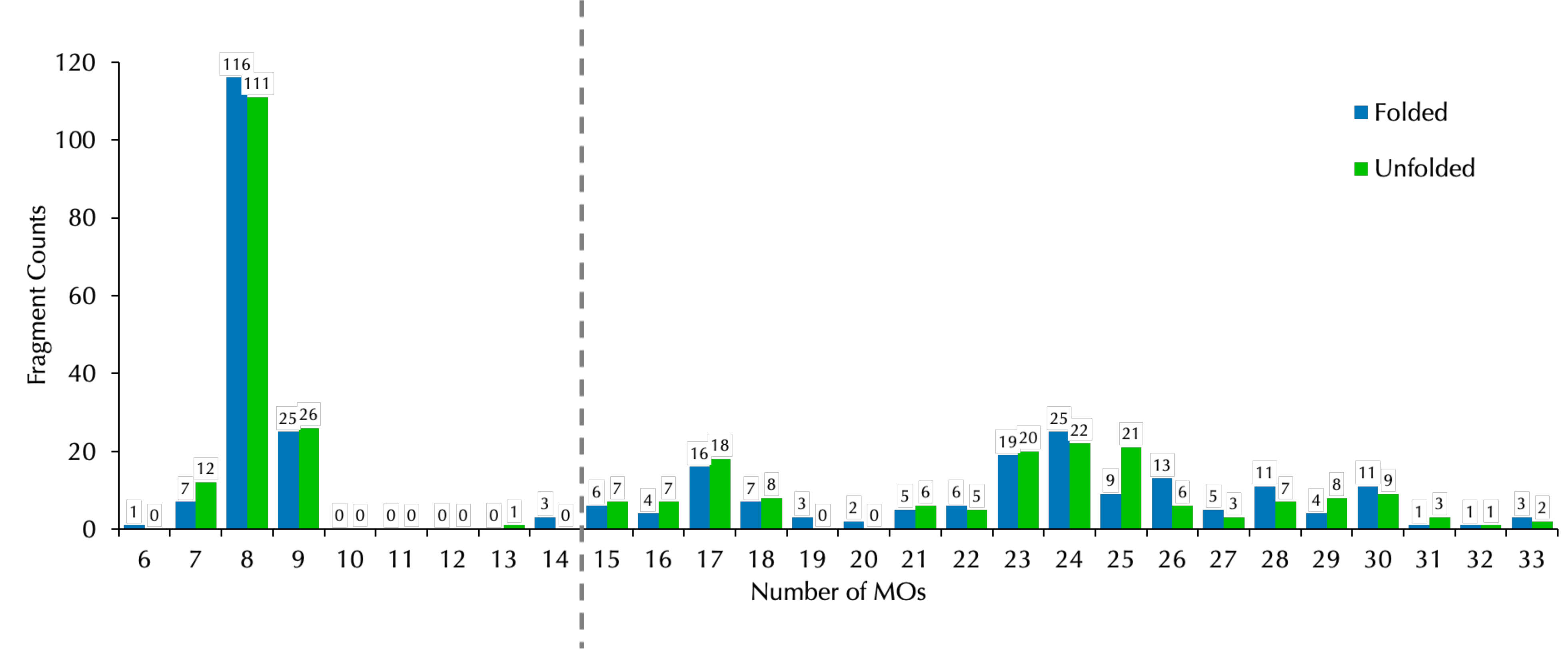}
    \caption{Number of EWF clusters with a given size (defined as the number of MOs) for the folded (blue) and unfolded (green) conformers. The black dashed line separates EWF clusters studied with FCI (left) and SQD (right).}
    \label{fig:number of fragments}
\end{figure}

Once all EWF clusters are generated and their corresponding Hamiltonians are saved as HDF5 files, the subsequent calculations are organized by grouping fragments according to their size, i.e. number of MOs. Each EWF cluster problem is then solved at a correlated level to obtain energies and reduced density matrices for the reconstruction step. Figure \ref{fig:number of fragments} illustrates the distribution of EWF cluster counts based on the number of molecular orbitals for two distinct molecular conformers: folded (blue bars) and unfolded (green bars). Overall, the EWF cluster counts for the folded and unfolded conformers are highly similar across most MO ranges. Both conformers exhibit a sharp maximum at the 8-MO bin, where the folded conformer contains 116 EWF clusters and the unfolded conformer contains 111 EWF clusters. These EWF clusters are almost exclusively hydrogen-atom environments and are efficiently solved using classical full configuration interaction (FCI).

Beyond 14 MOs, the EWF cluster counts become lower and more sparse. These larger EWF clusters (15–33 MOs) correspond to heavy-atom fragments involving carbon, nitrogen, and oxygen. For both conformers, the largest EWF clusters contain 33 MOs. At this scale, the computational cost of exact diagonalization becomes prohibitive. To treat these cases efficiently, EWF clusters in the 15–33 MO range are solved using sample-based quantum diagonalization (SQD), which uses quantum-device sampling and classical post-processing to approximate the ground-state energy and reduced density matrices.

While the small-MO EWF clusters ($<$15 MOs) are trivial to solve with FCI, the SQD calculations for larger EWF clusters require substantially more computational time and memory, with costs increasing steeply as the MO count grows. To manage these demands, the post-processing of quantum samples and SQD evaluations is distributed across multiple HPC resources. Fragments in the 15–24 MO range are assigned to lighter HPC nodes, whereas the 25–33 MO EWF clusters, which are the most computationally intensive, are handled on heavy HPC resources to ensure feasible turnaround times.

The lighter HPC resources correspond to the MSU (Michigan State University) HPC, where SQD post-processing for the 15–24 MO EWF clusters are run in parallel across 80–100 nodes simultaneously, each providing 493 GB of RAM and 128 CPU cores, with 48 cores per node allocated for the computations. The most demanding 25–33 MO EWF clusters are routed to heavy HPC resources on the CCF (Cleveland Clinic Foundation) HPC, which include multiple 3 TB nodes with 96 CPUs and several 6 TB nodes with 112 CPUs. All SQD post-processing tasks consistently use 48 CPUs per node, while the RAM footprint scales with the molecular orbital size of each EWF cluster, ensuring efficient resource utilization across the full spectrum of EWF cluster sizes.

\subsubsection{Quantum circuit design: the LUCJ ansatz}

We start from the active-space Hamiltonian, written in second quantization as
\begin{equation}\label{eq3}
    \hat{H} = E_0 + \sum_{\substack{pr \\ \sigma}}h_{pr}\hat{a}_{p\sigma}^\dagger \hat{a}^{\phantom{\dagger}}_{r\sigma}+\sum_{\substack{prqs \\ \sigma\tau}} \frac{(pr|qs)}{2} \hat{a}_{p\sigma}^\dagger \hat{a}_{q\tau}^\dagger \hat{a}^{\phantom{\dagger}}_{s\tau} \hat{a}^{\phantom{\dagger}}_{r\sigma}
    \;,
\end{equation}
where $\hat{a}^\dagger$ ($\hat{a}$) are creation (annihilation) operators, $p$,$r$,$s$, and $q = 1 \dots M$ denote basis set elements (MOs), $\sigma$ and $\tau$ denote spin-$z$ polarizations, $h_{pr}$ and $(pr|qs)$ are the one- and two-body electronic integrals, and $E_0$ is a constant accounting for the electrostatic interactions between nuclei and electrons in occupied inactive orbitals. We obtain the quantities $E_0$, $h_{pr}$, and $(pr|qs)$ for the selected active spaces using PySCF.\cite{sun2020recent,sun2018pyscf,sun2015libcint}

We prepare our quantum circuits, used to sample configurations, starting from a truncated version of the local unitary cluster Jastrow (LUCJ) ansatz~\cite{motta2023bridging}
\begin{equation}
\label{eq4}
|\Psi\rangle=\prod_{\mu=0}^{L-1}e^{\hat{K}_\mu}e^{i\hat{J}_{\mu}}e^{-\hat{K}_\mu} 
| {\bf{x}}_{\mathrm{RHF}} \rangle \;,
\end{equation}
where ${\hat{K}_\mu}=\Sigma _{pr,\sigma} K^{\mu}_{pr}\hat{a}_{p\sigma}^\dagger\hat{a}^{\phantom{\dagger}}_{r\sigma}$ are one-body operators, ${\hat{J}_\mu}=\Sigma _{pr,\sigma \tau} J^{\mu}_{p\sigma, r\tau}\hat{n}_{p\sigma}\hat{n}_{r\tau}$ are density-density operators, and $| {\bf{x}}_{\mathrm{RHF}} \rangle$ is the restricted closed-shell Hartree-Fock (RHF) state. We use the Jordan-Wigner (JW) transformation~\cite{ortiz2002simulating} to map the fermionic wavefunction Eq.~\eqref{eq4} onto a qubit wavefunction that can be prepared executing a quantum circuit. 
The JW transformation maps the Fock space of fermions in $M$ spatial orbitals onto the Hilbert space of $2M$ qubits, where the basis state $|{\bf{x}} \rangle$ is parametrized by a bitstring ${\bf{x}} \in \{0,1\}^{2M}$ and represents an electronic configuration where the spin-orbital $p\sigma$ is occupied (empty) if $x_{p\sigma}=1$ ($x_{p\sigma}=0$). 
We prepare the wavefunction given in Eq.~\eqref{eq4} by executing the following quantum circuit: a single layer of Pauli-X gates prepares the basis state $| {\bf{x}}_{\mathrm{RHF}} \rangle$, a Bogoliubov circuit~\cite{aleksandrowicz2019qiskit} (with linear depth, quadratic number of gates, and a 1D qubit connectivity) encodes each orbital rotation $e^{ \pm \hat{K}_\mu}$, and a circuit of Pauli-ZZ rotations encodes each density-density interaction $e^{i\hat{J}_{\mu}}$. When $J^{\mu}$ is a dense matrix, Pauli-ZZ rotations are applied across all pair of qubits, requiring all-to-all qubit connectivity or a substantial overhead of SWAP gates. To mitigate these quantum hardware requirements LUCJ imposes a ``locality'' approximation, e.g., in the most extreme form it enforces $J^{\mu}_{p\sigma, r\tau}=0$ for all pairs of spin-orbitals that are not mapped onto adjacent qubits under JW~\cite{motta2023bridging} (as a consequence, a circuit with constant depth and linear number of gates encodes each $e^{i\hat{J}_{\mu}}$ operator). Furthermore, to control the depth and size of quantum circuits, the number of layers in Eq.~\eqref{eq4} is $L=1$.
We produce the LUCJ circuits using the ffsim library~\cite{ffsim2024} interfaced with Qiskit 1.1.1~\cite{aleksandrowicz2019qiskit,javadi2024quantum}.

\textbf{Initializing LUCJ circuits with CCSD amplitudes:}
Classical CCSD (coupled cluster with singles and doubles) approximates the ground-state wavefunction using an exponential ansatz,
\begin{align}
    \ket{\Psi_{\mathrm{CCSD}}} = e^{\hat{T}_1+\hat{T}_2}\ket{\Psi_0}
\end{align}
where

\begin{align}
\hat{T}_1 = \sum_{ai} t^a_i \, \hat{a}^\dagger_a \hat{a}_i
\quad,\quad 
\hat{T}_2 = \sum_{aibj} t^{ab}_{ij} \, \hat{a}^\dagger_a \hat{a}^\dagger_b \hat{a}_j \hat{a}_i
\quad.
\end{align}

LUCJ circuits can be initialized using CCSD t-amplitudes via a double-factorized representation~\cite{motta2021low} of the $\hat{T}_2$ operators. We rewrite $\hat{T}_2-\hat{T}_2^\dagger$ in unitary CCSD theory as
\begin{align}
    \hat{T}_2-\hat{T}_2^\dagger = i\sum_{\mu=0}^{L-1} \hat{U}_\mu \hat{J}_\mu \hat{U}_\mu^\dagger \quad,
\end{align}
where $\hat{U}_\mu$ are orbital rotations and $\hat{J}_\mu$ are diagonal Coulomb operators.

Applying a Trotter approximation,
\begin{align}
    e^{\hat{T}_2-\hat{T}_2^\dagger}\approx \sum_{\mu=0}^{L-1} \hat{U}_\mu e^{i \hat{J}_\mu}
    \hat{U}_\mu^\dagger.
\label{eq:ucj_lr}
\end{align}
Thus, CCSD $t_2$-amplitudes provide an initial guess for UCJ via double factorization. To make Eq.~\eqref{eq:ucj_lr} compatible with LUCJ, we omit matrix elements of $\hat{J}_\mu$ that are inaccessible under qubit connectivity constraints.

\begin{figure}[h] 
    \centering 
    \includegraphics[width=0.8\textwidth]{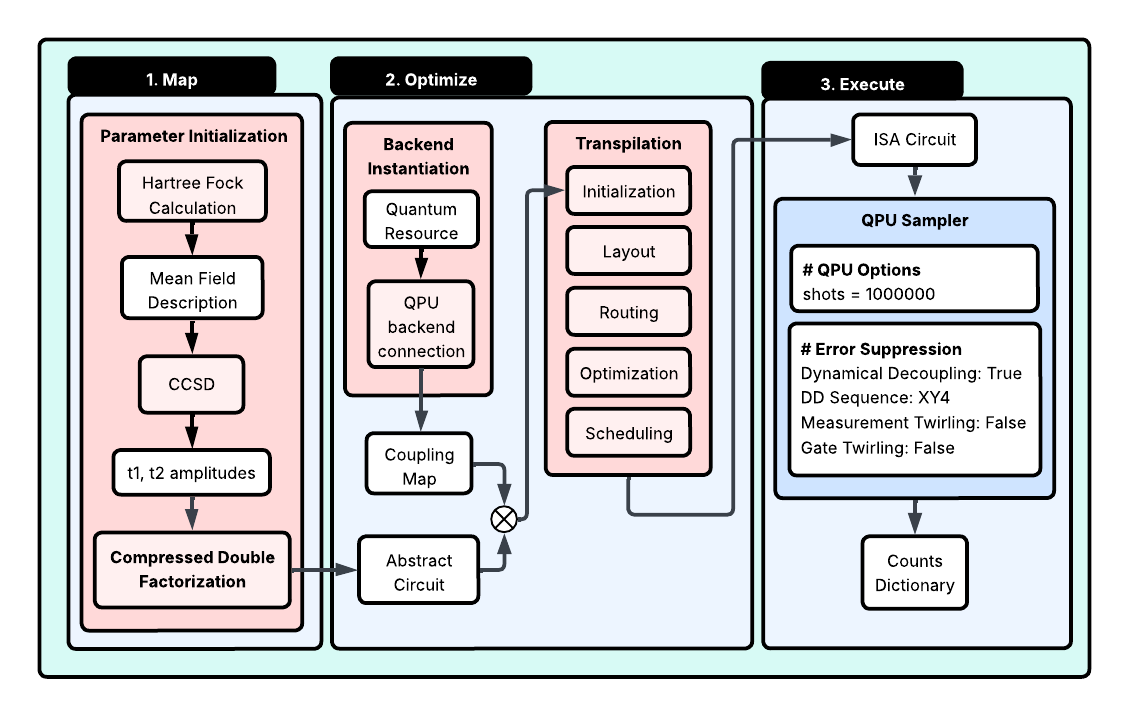} 
    \caption{Exploded schematic of the LUCJ sampling workflow used in SQD calculations. LUCJ circuits are initialized from Hartree--Fock and CCSD amplitudes via compressed double factorization, instantiated and transpiled for a target QPU backend, executed with error-suppression protocols, and iteratively optimized through classical post-processing of measurement outcomes.}
    \label{fig:lucj_sampling} 
\end{figure}

Figure~\ref{fig:lucj_sampling} illustrates the sampling and optimization workflow for LUCJ circuits, highlighting the interaction between circuit execution, sampling, and classical parameter updates.

\textbf{Parameter Optimization:}
We optimize the LUCJ parameters according to the guidelines in Lin et al~\cite{lin2025pushing}. A recent study showed significant improvements of the sampling ability upon optimization of the LUCJ parameters \cite{Lin_2025_ImprovedParameterInitialization}.
The performance of the LUCJ circuits is improved by applying a compressed double factorization to the $t_2$-amplitudes of $\hat{T}_2$ operator:
\begin{align}
    \overline{t}_{ij}^{ab} = i\sum_{\mu=0}^{L-1}\sum_{pq} J_{pq}^{(\mu)} U_{ap}^{(\mu)} U_{ip}^{(\mu)*} U_{bq}^{(\mu)} U_{jq}^{(\mu)*},
\label{eq:tbar}
\end{align}
where each $U^{(\mu)}$ is an orbital rotation and each $J^{(\mu)}$ is a Coulomb matrix. Here, the truncated $U^{(\mu)}$ and $J^{(\mu)}$ are optimized to minimize
\begin{align}
    \chi = \frac{1}{2}\sum_{ijab}|\bar{t}_{ij}^{ab}-t_{ij}^{ab}|^2,
\end{align}
where $\bar{t}_{ij}^{ab}$ are as in Eq.~\eqref{eq:tbar} and $t_{ij}^{ab}$ are the original CCSD $t_2$-amplitudes.
Optimizations employ L-BFGS-B in SciPy with gradients obtained using automatic differentiation. In the present study we use single-stage optimization due to the fact that multi-stage optimization \cite{Lin_2025_ImprovedParameterInitialization} is computationally expensive for the largest EWF-cluster.

\subsubsection{Sampling using circuits run on quantum hardware}

\begin{figure}
    \centering
    \includegraphics[width=0.5\linewidth]{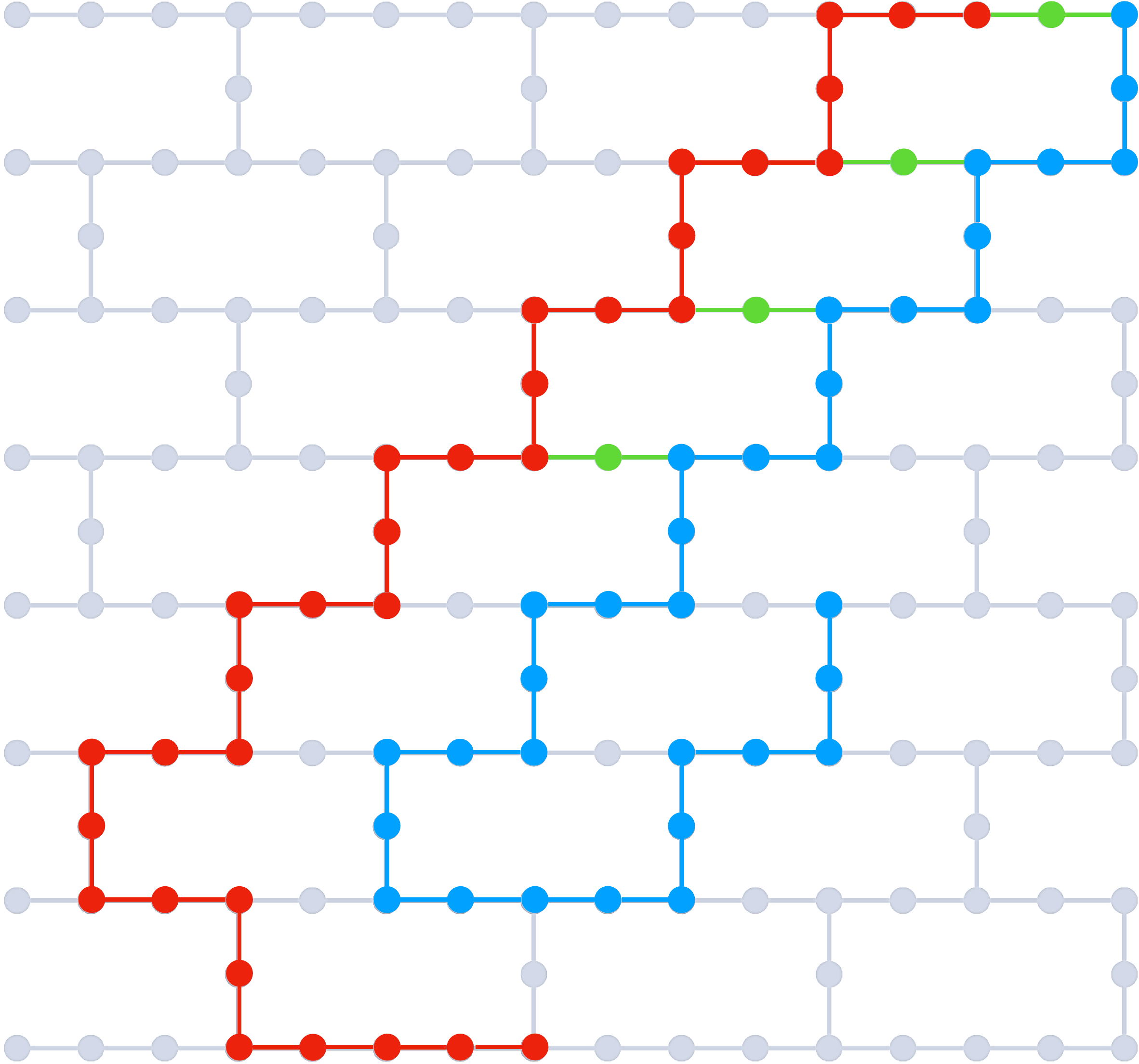}
    \caption{Qubit layout for the largest EWF cluster with 33 molecular orbitals. Red and blue dots indicate $\alpha$- and $\beta$-qubits, respectively, and the four green dots are ancillary qubits connecting four 
    $\alpha$–$\beta$ pairs.}
    \label{fig:qubit_layout}
\end{figure}

For the hardware simulations, all circuits were executed on IBM Heron-R2 QPUs with 1,000,000 measurement outcomes (shot) per circuit. LUCJ circuits with $L=1$ were used and four $\alpha$–$\beta$ qubits were connected via ancillary qubits (see Fig. \ref{fig:qubit_layout} for the qubit layout). Among the available zig-zag connectivity patterns, the layout was selected using a heuristic that sums the two-qubit gate errors of all couplers and the readout errors of the qubits in the pattern~\cite{IBM2025SQD}.

Qiskit v2.1.1\cite{javadi2024quantum} was used with $generate\_preset\_pass\_manager$ at $optimization\_level=0$, incorporating ffsim's $PRE\_INIT$ passes for orbital-rotation decomposition and applying $RemoveIdentityEquivalent$ post-initialization to remove negligible (i.e. near-identity) rotations. Dynamic decoupling using the $XY4$ sequence was applied to further suppress decoherence.

\begin{figure}[htp]
    \centering
    \includegraphics[width=1\linewidth]{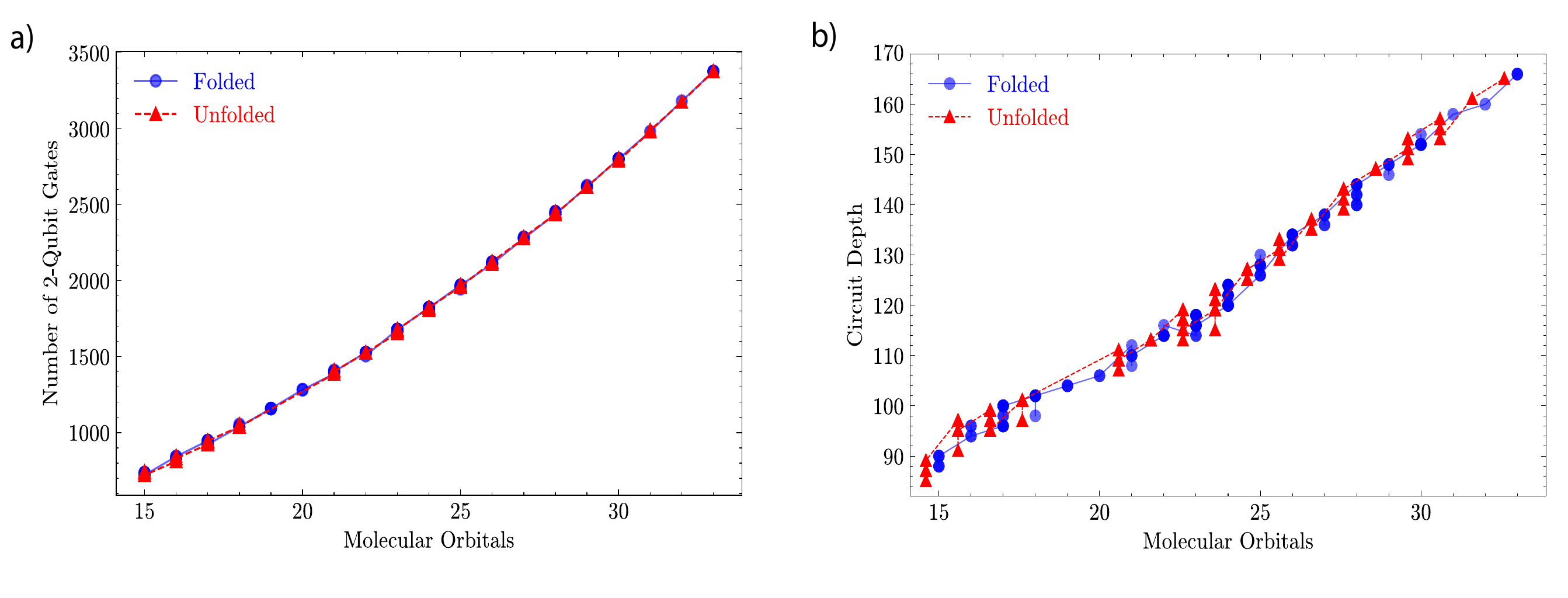}
    \caption{Quantum circuit complexity across the EWF-SQD fragments expressed through: A) CNOT gate count; B) 2-qubit gate depth.}
    \label{fig:circuit_chars}
\end{figure}

\textbf{Quantum resources}
Figure \ref{fig:circuit_chars} illustrates the relationship between the number of molecular orbitals and two key characteristics of a quantum circuit: the number of 2-qubit gates (a) and the circuit depth (b). Both characteristics are analyzed for two different conformers: folded (represented by blue circles) and unfolded (represented by red triangles). 

Panel (a) shows the required number of 2-qubit gates increases quadratically with the number of molecular orbitals, ranging from approximately 750 at 15 orbitals up to around 3400 at 32 orbitals. Panel (b) shows that the circuit depth increases linearly from 90 to 160 with the number of molecular orbitals.

\subsubsection{Ground State and Energy of the Hamiltonian in SQD}

In this section, we discuss the SQD calculations of the ground state and energy of the Hamiltonian for the systems studied. We describe how the initially collected candidate electron configurations from the quantum device undergo the configuration recovery procedure and diagonalization performed on each step of configuration recovery, as well as the augmentation of the electron configuration susbpace performed in the extended SQD (ext-SQD) procedure.

\textbf{Configuration recovery:}
Upon executing the LUCJ circuits, we measure $|\Psi \rangle$ in the standard computational basis. Repeating this produces a set of measurement outcomes (or ``shots'') 
\begin{equation}\label{eq5}
\tilde{\chi} = \left \{ {\bf{x}} | {\bf{x}} \sim  \tilde{p}({\bf{x}}) \right \}
\end{equation}
in the form of bitstrings ${\bf{x}} \in \left \{0,1 \right \}^{2M}$, each representing an electronic configuration (Slater determinants) distributed according to $\tilde{p}({\bf{x}})$.
While in a noiseless device the configurations are distributed according to $| \langle {\bf{x}} | \Psi \rangle|^2$, on a noisy device they follow a distribution $\tilde{p}({\bf{x}}) \neq | \langle {\bf{x}} | \Psi \rangle|^2$. In particular, $\tilde{p}({\bf{x}})$ breaks particle number conservation and returns configurations with incorrect particle number. 

We use a technique called self-consistent configuration recovery~\cite{robledo2024chemistry}, executed on a classical computer, to restore particle number conservation. The associated code is publicly available in the GitHub repository.~\cite{sqd_addon}
Within each step of self-consistent recovery, we sample $K$ subsets (or batches) of $\tilde{\chi}$ labeled $\tilde{\chi}_b$ with $b = 1 \dots K$. Each batch defines -- through a transformation~\cite{robledo2024chemistry} informed by an approximation to the ground-state occupation numbers $n_{p\sigma}$ -- a subspace $S^{(b)}$ of dimension $d$, in which we project the many-electron Hamiltonian as~\cite{kanno2023quantum,nakagawa2023adapt,robledo2024chemistry}
\begin{equation}
\label{eq6}
\hat{H}_{S^{(b)}}=\hat{P}_{S^{(b)}}\hat{H}\hat{P}_{S^{(b)}}
\;,
\end{equation}
where the projector $\hat{P}_{S^{(b)}}$ is
\begin{equation}
\label{eq7}
\hat{P}_{S^{(b)}} =\sum_{ {\bf{x}} \in S^{(b)}} | {\bf{x}} \rangle \langle {\bf{x}} |
\;.
\end{equation}
In this work, we use 3000 samples per batch and 10 batches for all of the EWF clusters. We compute the ground states and energies of the Hamiltonians in Eq.~\eqref{eq6}, $|\psi^{(b)} \rangle$ and $E^{(b)}$ respectively, and use the lowest energy across the batches, $\min_b E^{(b)}$, as the best approximation to the ground-state energy at the current iteration of the configuration recovery. We use the ground states $|\psi^{(b)} \rangle$ to obtain an updated set of occupation numbers,
\begin{equation}
\label{eq8}
n_{p\sigma}=\frac{1}{K} \sum_{1\leq b\leq K } \langle \psi^{(b)} | \hat{n}_{p\sigma} |\psi^{(b)} \rangle
\;,
\end{equation}
that we use in the next iteration of configuration recovery to produce the subspaces $S^{(b)}$. We repeat the iterations of self-consistent configuration recovery until convergence of the energy $\min_b E^{(b)}$, average orbital occupancy $\hat{n}_{p\sigma}$, or until we reach the maximum number of iterations. 

Moreover, we set the convergence thresholds for total energy and average orbital occupancy as $\scientific{1}{-8}$ and $\scientific{1}{-5}$ respectively, and we set the maximum number of configuration recovery iterations at 5. We use ``Qiskit Addon: SQD Version 0.12.0'' in which we select the most relevant electron configurations, based on their configuration interaction coefficients at each step of the configuration recovery procedure. These selected configurations are always included in all the batches produced. \cite{Shirakawa_2025_ClosedLoopQuantumClassical} The threshold for identifying configurations as relevant was set to $\scientific{1}{-4}$. All simulations were performed with the spin symmetrization for $\alpha$ and $\beta$ spin-orbitals. In the first iteration of self-consistent configuration recovery, we initialize $n_{p\sigma}$ from  measurement results in $\tilde{\chi}$ with the correct particle number.

\begin{figure}[h] 
    \centering 
    \includegraphics[width=0.8\textwidth]{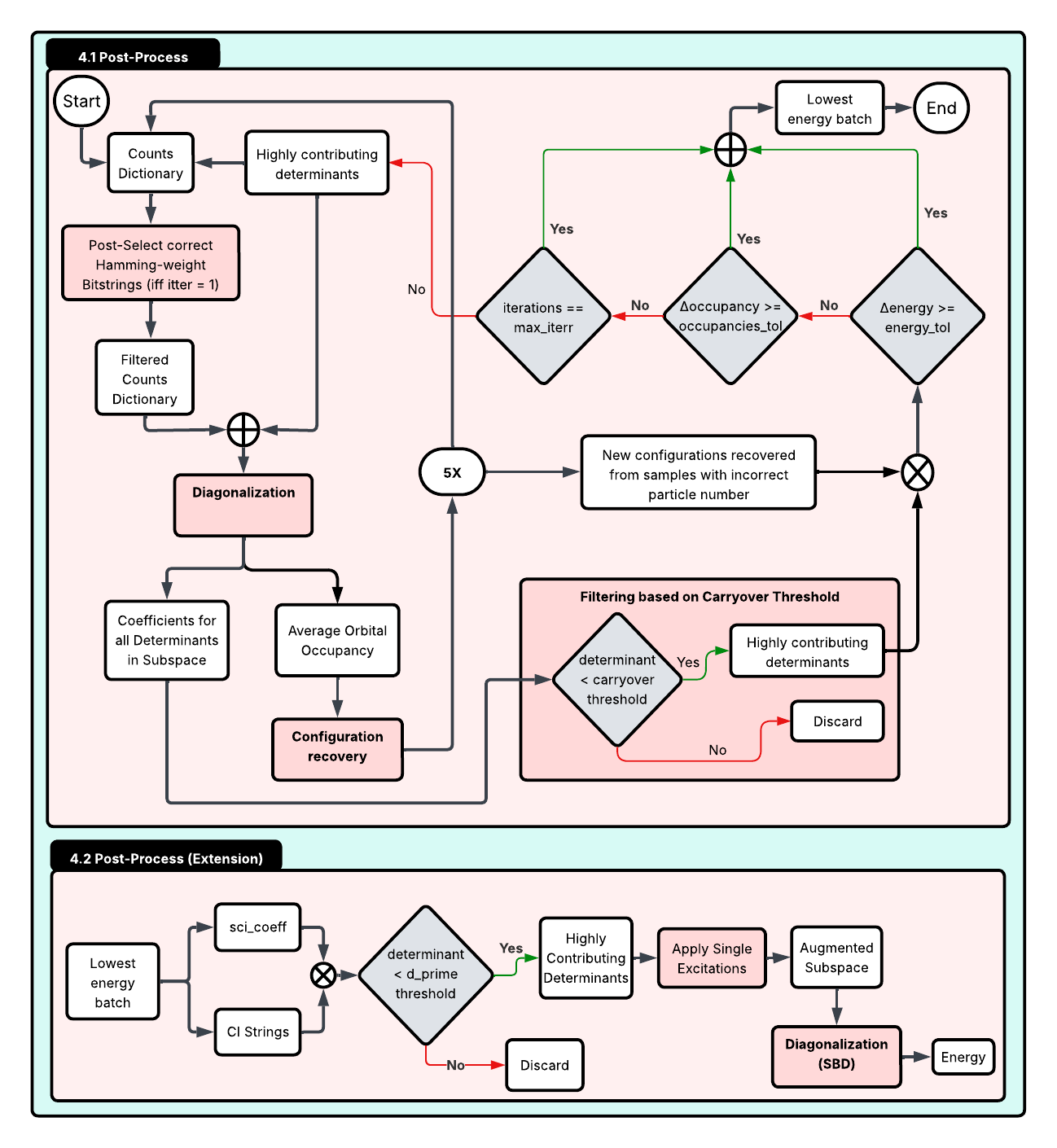} 
    \caption{Exploded schematic of the SQD classical post-processing pipeline. Measurement outcomes are filtered and post-selected to restore particle-number conservation, followed by batch-wise construction and diagonalization of reduced configuration subspaces. Convergence is assessed using energy and orbital-occupancy criteria, and in extended SQD (ext-SQD) the dominant determinants are augmented with single excitations to improve ground-state and excited-state accuracy.}
    \label{fig:sqd_exploded} 
\end{figure}

The diagonalization in each step of configuration recovery was performed with the SBD solver which can be obtained through the GitHub repository.\cite{sbd} The diagonalization in the SBD solver is parallelized with 48 CPUs used in the diagonalization of each subspace, where we use a total of 480 CPUs in 10 batches for each EWF cluster. The full sequence of configuration recovery, subspace construction, and diagonalization steps is summarized schematically in Fig.~\ref{fig:sqd_exploded}.

\textbf{Subspace augmentation (ext-SQD):}
The extended SQD (ext-SQD) procedure was introduced to enable excited-state calculations based on SQD~\cite{barison2025quantum,shivpuje2025sample}
but may also be used to improve the accuracy of ground- and excited-state SQD calculations~\cite{barroca2025surface}.

In ext-SQD we first take the electron configurations from the lowest energy batch on the last step of the configuration recovery procedure. We select the most dominant electron configurations based on the value of their configuration interaction coefficients (dominant if the threshold is above $\scientific{1}{-5}$). Using the selected configurations, we then augment their subspaces by applying single excitations on each configuration within the selected set. The augmentation of the subspace was peformed with the PyCI software package.\cite{richer2024pyci} We use the augmented subspace to perform diagonalization within the given EWF cluster to obtain the ground state and energy of the Hamiltonian. Finally, we use the ground state produced after the diagonalization of the augmented subspace to calculate the reduced density matrices (RDM) of each EWF cluster.

\subsubsection{Classical complexity}

\begin{figure} [h]
    \centering
    \includegraphics[width=0.8\linewidth]{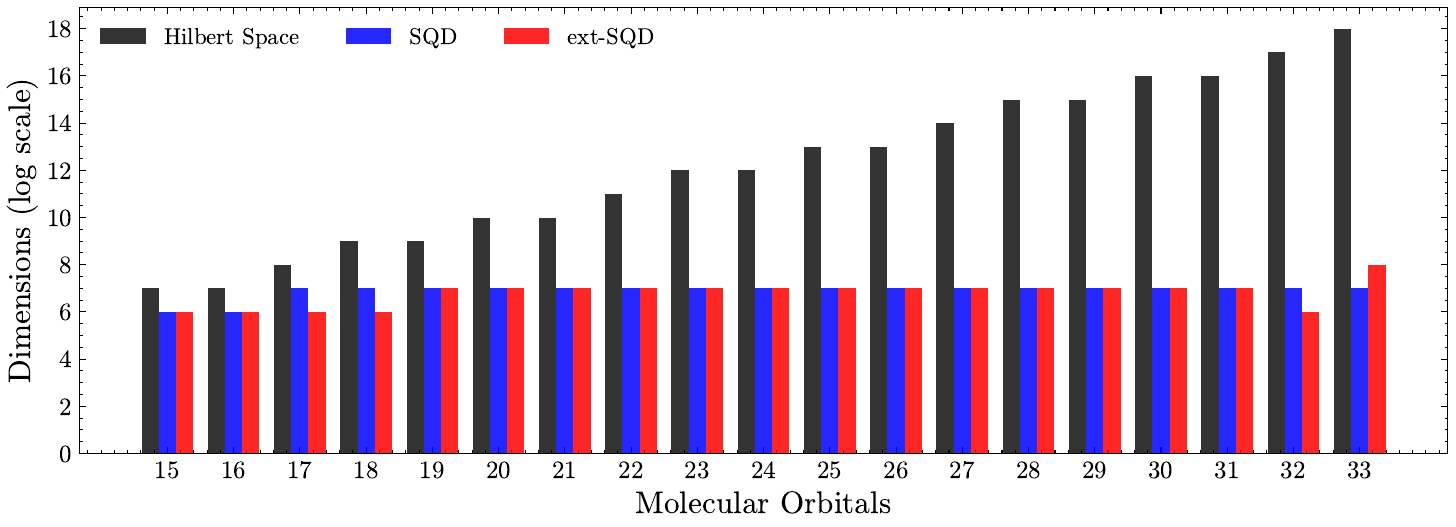}
    \caption{The average dimensions of full Hilbert space, SQD subspace, and ext-SQD subspace across EWF clusters for the given number of molecular orbitals. The dimensions of subspaces shown in the logarithmic scale.}
    \label{fig:subspace-dims}
\end{figure}

Figure~\ref{fig:subspace-dims} shows that the use of configuration sampling produced with LUCJ and the configuration recovery procedure reduces the effective SQD and ext-SQD subspaces by two orders of magnitude for small fragments and by more than ten orders of magnitude for larger fragments compared to the full Hilbert space.

\begin{figure}[h]
    \centering
    \includegraphics[width=0.75\linewidth]{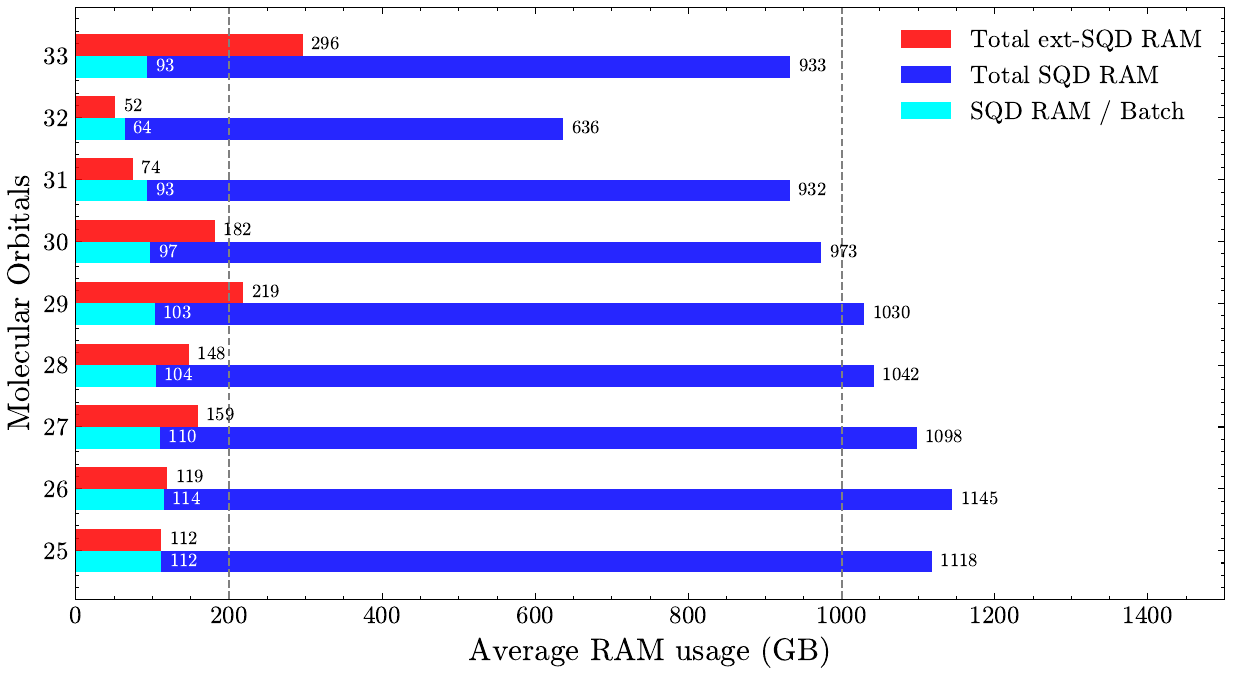}
    \caption{The average memory requirements of the SQD and ext-SQD steps of the workflow for EWF clusters with the given number of molecular orbitals ranging between 25-33. Dashed vertical line marks the 200 GB and 1000GB.}
    \label{fig:memory}
\end{figure}

Figure \ref{fig:memory} shows that even though individual batches of SQD simulations use less RAM than ext-SQD simulations, the total average usage of RAM in the SQD step exceeds the total amount of the average RAM used in the ext-SQD step due to the usage of multiple batches in the SQD steps. Hence, ext-SQD allows better memory efficiency in the post processing step by utilizing only the single lowest energy batch. 

\subsection{Reconstruction of Global Energies from EWF Clusters}
After all EWF cluster calculations are completed, the final step is to assemble the global energy of the full Trp-cage conformer from the independent EWF cluster results. This is achieved through a collating procedure based on global expectation values derived from the EWF cluster density matrices. From each EWF cluster solver, whether FCI or SQD, the one- and two-body reduced density matrices in the embedded orbital space are extracted. These matrices encode both the fragment-specific correlation and the correlation-driven couplings between each fragment and its bath.

The collating procedure constructs the global one- and two-body expectation values by projecting each EWF cluster density matrix back into the full system and symmetrizing contributions over overlapping bath spaces. Since fragment orbitals are non-overlapping while bath orbitals generally overlap, this symmetrization is essential to avoid double counting while retaining all physical contributions. The resulting global expectation values yield an N-representable set of observables that can be used to compute the total embedded energy of each Trp-cage conformer. Finally, the relative energy between the folded and unfolded conformers is obtained directly from these global energies.

\section{Results}

Table~\ref{table:ewf_sqd} summarizes the relative energies of the folded and unfolded Trp-cage conformers obtained using different EWF-based electronic structure methods. By comparing EWF-MP2, EWF-CCSD, and EWF-(FCI,SQD), we assess the impact of both correlation level and the use of SQD as a fragment solver within a fixed EWF fragmentation scheme.

\begin{table}
\begin{tabular}{rlll}
\hline\hline
method & $E_{\mathrm{unfolded}} \, [E_{\mathrm{h}}]$ & $E_{\mathrm{folded}} \, [E_{\mathrm{h}}]$ & $\Delta E$ [kcal/mol]\\
\hline
EWF-MP2       & -7354.0653 & -7354.1413 & 59.61 \\
EWF-CCSD      & -7354.0653 & -7354.1413 & 47.70 \\
EWF-(FCI,SQD) & -7354.1372 & -7354.2256 & 55.43 \\
\hline\hline
\end{tabular}
\caption{Ground-state energies of the Trp-cage in the unfolded and folded conformations (second and third columns) and the corresponding relative energies (fourth column) obtained using EWF-based methods.}
\label{table:ewf_sqd}
\end{table}

At the EWF-MP2 level, the folded structure is stabilized relative to the unfolded conformer by 59.61~kcal/mol. Inclusion of higher-order correlation through EWF-CCSD significantly reduces this stabilization to 47.70~kcal/mol.

Using SQD as a fragment solver within the EWF framework yields a relative energy of 55.43~kcal/mol. The difference of -7.7~kcal/mol with respect to EWF-CCSD and 4.2 ~kcal/mol with respect to EWF-MP2 reflects the combined effect of SQD subspace selection and the approximate treatment of larger fragments, while maintaining the same fragmentation protocol across methods.

\section{Conclusions and Outlook}

This study presents two important breakthroughs. First, the simulation of Trp-cage with STO-3G basis set is the largest example of embedding-based simulations within the configuration interaction paradigm and to our knowledge this is the largest case ever simulated exclusively within the configuration interaction framework. Trp-cage simulated with the STO-3G basis set results in simulation of 919 molecular orbitals (MOs), where the largest DMET and EWF simulations were reported only for systems explicitly including only up to 321 and 312 MOs, respectively.\cite{wouters2016practical, nusspickel2022systematic} Second, this represents the first demonstration of a system of this biochemical size being treated with quantum sampling on a quantum device as part of an embedding workflow, illustrating both the feasibility and the growing capability of hybrid quantum–classical electronic structure methods. 

The present study paves the way for fragment-based solutions in quantum computing. Our approach is highly modular and allows for integration of novel quantum computing methodologies in the future. Hence, the present workflow will gradually mature in a synchronous manner with the maturation of quantum hardware and quantum algorithms. In the long-term, upon establishment of fault-tolerant quantum computing, the present fragment-based methodology can eventually adapt even more powerful fully-quantum methods such as quantum phase estimation (QPE) where perspectives of utilization of this method within embedding schemes were recently explored by Erakovic et al.\cite{erakovic2025high}

Configuration interaction simulations accelerated with quantum computers can offer unique opportunities for simulations of complex chemical processes in proteins such as photochemical reactions and proton-coupled electron transfer\cite{mai2020molecular,warburton2022theoretical}. In the long-term, upon adaptation of more advanced future quantum hardware and  quantum algorithms, this methodology potentially might offer a unified highly-accurate method for the treatment of any protein-based system beyond the scaling of classical computers. 

\section*{Acknowledgments}

The authors gratefully acknowledge financial support from the National Science Foundation (NSF) through CSSI Frameworks Grant OAC-2209717 and from the National Institutes of Health (Grant Numbers GM130641). The authors are grateful to the high-performance computer center (iCER HPCC) at Michigan State University and the high-performance computer center at Cleveland Clinic Foundation. We thank Abdullah Ash Saki for providing code to pick good qubit layouts for the LUCJ circuits on IBM QPUs. We thank George H. Booth for helpful discussion on the EWF method which was essential for scaling of the simulations within this method. We thank Milana Bazayeva and Zhen Li for their helpful feedback during the development of software components,  which were necessary for integration of the SBD solver within the EWF-(FCI,SQD) workflow.

\newpage

\bibliography{achemso-demo}

\end{document}